\documentclass[copyright,creativecommons]{eptcs}
\providecommand{\event}{Festschrift DS 2013} %
\usepackage{breakurl}             %
\usepackage{alltt}
\usepackage{amsmath}
\usepackage{amssymb}
\usepackage{color}

\usepackage{cite}

\newcommand{\submit}[1]{}

\newcommand{\ignore}[1]{}              %
\newcommand{\hidecomment}[1]{}         %

\newcommand{\eg}{{\it e.g.}}

\newcommand{\cf}{{\it cf.}}
\newcommand{\etal}{{\it et al.}}

\newcommand{\fontAcode}[1]{\textsf{#1}}
\newcommand{\fc}[1]{\fontAcode{#1}}  %

\newcommand{\PDA}{{PDA}}

\newcommand{\iDPDA}{{1DPDA}}
\newcommand{\iiDPDA}{{2DPDA}}
\newcommand{\iNPDA}{{1NPDA}}
\newcommand{\iiNPDA}{{2NPDA}}

\newcommand{\rawDint}{Int}
\newcommand{\rawDsim}{Sim}
\newcommand{\rawNint}{Int}
\newcommand{\rawNsim}{Sim}
\newcommand{\rawApush}{push}
\newcommand{\rawAop}{op}
\newcommand{\rawApop}{pop}
\newcommand{\rawAchoose}{choose}
\newcommand{\rawAaccept}{accept}
\newcommand{\rawAacceptop}{\textbf{accept}}
\newcommand{\rawAhalt}{halt}
\newcommand{\rawAhaltop}{\textbf{halt}}
\newcommand{\rawAdefined}{defined}
\newcommand{\rawAvisited}{visited}
\newcommand{\rawAnext}{next}
\newcommand{\rawAnextl}{nextleft}  %
\newcommand{\rawAnextr}{nextright}
\newcommand{\rawAfollow}{follow}

\newcommand{\Dint}[1]{\rawDint({#1})}
\newcommand{\Dsim}[1]{\rawDsim({#1})}
\newcommand{\Nint}[1]{\rawNint({#1})}
\newcommand{\Nsim}[1]{\rawNsim({#1})}
\newcommand{\Apush}[1]{\rawApush({#1})}
\newcommand{\Aop}[1]{\rawAop({#1})}
\newcommand{\Apop}[1]{\rawApop({#1})}
\newcommand{\Achoose}[1]{\rawAchoose({#1})}
\newcommand{\Aaccept}[1]{\rawAaccept({#1})}
\newcommand{\Aacceptop}{\rawAacceptop}
\newcommand{\Ahalt}[1]{\rawAhalt({#1})}
\newcommand{\Ahaltop}{\rawAhaltop}
\newcommand{\Adefined}[1]{\rawAdefined({#1})}
\newcommand{\Avisited}[1]{\rawAvisited({#1})}
\newcommand{\Anext}[1]{\rawAnext({#1})}
\newcommand{\Anextl}[1]{\rawAnextl({#1})}
\newcommand{\Anextr}[1]{\rawAnextr({#1})}
\newcommand{\Afollow}[2]{\rawAfollow({#1},{#2})}

\newcommand{\myma}[1]{\ensuremath{#1}}

\newcommand{\myamp}[0]{&}

\makeatletter
\def\framedstuffsidemargin{1em}
\def\framedstufftopmargin{.7em}

\makeatother

\newcommand{\mycaption}[1]{\caption{#1}}

\newcommand{\mypara}[1]{\paragraph{#1.}}
\newenvironment{codesize}{}{}

\title{Simulation of Two-Way
Pushdown Automata Revisited}
\author{Robert Gl\"uck
\institute{DIKU, Dept.\ of Computer Science, University of Copenhagen}
\email{glueck@acm.org}
}
\def\titlerunning{Simulation of Two-Way 
Pushdown Automata Revisited}
\def\authorrunning{R.\ Gl\"uck}

\begin{document}
\maketitle

\begin{center}
{\normalsize\it
Dedicated to David A.\ Schmidt on the Occasion of his 60th Birthday}
\end{center}

\begin{abstract}
The linear-time simulation of \emph{2-way deterministic pushdown automata} (\iiDPDA) by the Cook and Jones constructions is revisited. Following the semantics-based approach by Jones, an interpreter is given which, when extended with random-access memory, performs a linear-time simulation of \iiDPDA. The recursive interpreter works without the dump list of the original constructions, which makes Cook's insight into linear-time simulation of exponential-time automata more intuitive and the complexity argument clearer. The simulation is then extended to \emph{2-way nondeterministic pushdown automata} (\iiNPDA) to provide for a cubic-time recognition of context-free languages. The time required to run the final construction depends on the degree of nondeterminism. The key mechanism that enables the polynomial-time simulations is the sharing of computations by memoization.
\end{abstract}

\section{Introduction}
\label{sec:intro}

We revisit a result from theoretical computer science from a programming perspective. Cook's surprising theorem~\cite{Cook:72} showed that \emph{two-way  deterministic pushdown automata} (\iiDPDA) can be simulated faster 
on a random-access machine
(in linear time) than they may run natively (in exponential time). This insight %
was utilized by Knuth~\cite{KnuMorPra:77} to find a linear-time solution for the left-to-right
pattern-matching problem, which can easily be expressed as a {\iiDPDA}:
\begin{quote}
``\textit{This was the first time in Knuth's experience that automata theory had taught him how to solve a real programming problem better than he could solve it before.}''~\cite[p.~339]{KnuMorPra:77}
\end{quote}
Cook's original construction in 1971 is obscured by the fact that it does not follow the control flow of a pushdown automaton running on some input. It traces all possible flows backward thereby examining many unreachable computation paths, which makes the construction hard to follow.
Jones 
clarified the essence of the construction using a semantics-based simulator 
that interprets the automaton in linear time while following the control flow forward thereby avoiding unreachable branches~\cite{Jones:77}.
The simulator models the \emph{symbol stack} of the automaton on its \emph{call stack} using recursion in the meta-language and maintains
a local list of surface configurations (dump list) to record their common terminator in a table when a pop-operation is simulated.

We follow Jones' 
semantics-based
approach and give a simplified recursive simulator that does not require a local dump list and 
captures the essence of Cook's speedup theorem in a (hopefully) 
intuitive and easy to follow form. Furthermore, we then extend the construction from a
simulation of deterministic automata 
to a
simulation of \emph{two-way nondeterministic pushdown automata} ({\iiNPDA}).
The simulations are all realized by deterministic computation on a random-access machine.

Even though %
some time has passed since the theorem was originally stated, it continues to inspire studies in complexity theory and on the computational power 
of more practical programming paradigms, such as subclasses of imperative and functional languages (\eg~\cite{AT:92:TCS,AJ:94,Jones:97:complexity,Mogensen:94}).
It therefore appears worthwhile to capture the computational meaning of this classic result in clear and simple terms from a programming perspective. It is hoped that the progression from simple interpreters to simulators with memoization and termination detection makes these fundamental theoretical results more accessible.

We begin with a simple interpreter for two-way deterministic pushdown automata (Sect.~\ref{sec:Dint}) that we extend to simulate deterministic {\PDA} in linear time (Sect.~\ref{sec:Dsim}). We then introduce a nondeterministic choice operator (Sect.~\ref{sec:Nint}) and show the simulation of nondeterministic {\PDA} (Sect.~\ref{sec:Nsim}).

\section{Deterministic {\PDA} Interpreter}
\label{sec:Dint}

A \emph{two-way deterministic pushdown automaton} ({\iiDPDA}) consists of a finite-state control attached to a stack and an input tape with one two-way read-only head~\cite{Cook:72}. The state~$p$, the symbol read at head position~$i$, and the symbol~$A$ on top of the stack determine the next action for a given tape, which is the automaton's input. Only when the stack top is popped does the symbol below the top become relevant for the following computation. The set of states, the set of input symbols and the set of stack symbols are fixed for an automaton.
A transition function 
chooses the next action depending on the current \emph{surface configuration} $c=(p,i,A)$, shortly referred to as \emph{configuration}. 
The \emph{instantaneous description} 
$(c,\mbox{\it stack-rest})$ of an automaton includes the current configuration $c$ and the stack below the top symbol~$A$.

The automaton can {\bf push} and {\bf pop} stack symbols, and perform an operation {\bf op} that modifies the current configuration without pushing or popping (\eg, move to a new tape position).
The stack bottom and the left and right tape ends are marked by distinguished symbols.
The head position $i$ in a configuration $(p,i,A)$ is always kept within the tape bounds and one can determine an empty stack.
The automaton answers decision problems. 
It is said to \emph{accept} an input if, when started in initial state $p_0$ with an empty stack
and the head at the left endmarker, it terminates with {\bf accept}, an empty stack and the head at the right endmarker. It can just {\bf halt}
with an empty stack without accepting an input. In the exposition below we tacitly assume some fixed input tape.

\mypara{Termination}

A configuration in which a pop-operation occurs is a \emph{terminator}~\cite{Cook:72}. 
Every configuration~$c$ in a terminating computation has a unique terminator, that is the earliest terminator reached from $c$ that returns the stack \emph{below the height} at $c$. This case is illustrated below $(i)$: $d$ is the terminator of~$c$.
Terminator $d$ can be viewed as the result of configuration $c$. Configuration $c$ will always lead to $d$ regardless of what is on the stack below. A configuration that accepts or halts the automaton is also a terminator.

If a configuration $c$ is met again \emph{before} the terminator is reached, which means that the stack
never returned below the level at which $c$ occurred for the first time, then the automaton is in an \emph{infinite loop}. The second occurrence of $c$ will eventually lead to a third occurrence of $c$, ad infinitum.
The only two possible situations are illustrated below: either $c$ repeats at the same level of the stack $(ii)$ or at a higher level after some stack-operations have been performed~$(iii)$.
In both cases, the contents of the stack below $c$ (shaded) is untouched and irrelevant to the computation: $c$ will always lead to an infinite loop.
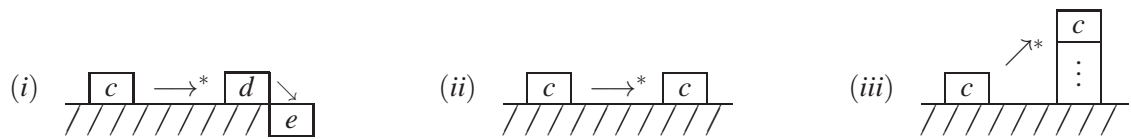
\newcommand{\myfbox}[1]{\framebox[6mm]{\rule[-0.7mm]{0mm}{3.3mm}{#1}}}
\newcommand{\myfboxtwo}[1]{\framebox[6mm]{\rule[-0.7mm]{0mm}{6.6mm}{#1}}}
\[
(i)~~~~~\begin{array}[t]{l@{~~\,}c@{~~}l}
    \myfbox{$c$}
  & \longrightarrow^*
  & \myfbox{$d$}\;\mbox{\scriptsize$\searrow$}\\
    \setlength{\unitlength}{1mm}
    \begin{picture}(0,0)
    \put(-3.1,3.62){
      \put(0,0){\line(1,0){27.1}}
      \multiput(0,-4)(3,0){9}{\line(1,2){2}}
      \put(27.03,-3.05){\myfbox{$e$}}}
  \end{picture}
\end{array}
\hspace{17mm}
(ii)~~~~~\begin{array}[t]{l@{~~\,}c@{~~}l}
    \myfbox{$c$}
  & \longrightarrow^*
  & \myfbox{$c$}\\
    \setlength{\unitlength}{1mm}
    \begin{picture}(0,0)
    \put(-3.1,3.62){
      \put(0,0){\line(1,0){30.5}}
      \multiput(0,-4)(3,0){10}{\line(1,2){2}}} 
  \end{picture}
\end{array}
\hspace{17mm}
(iii)~~~~~\begin{array}[t]{l@{~~}c@{~~}l}
  \myfbox{$c$}
  &
  \begin{minipage}[b]{5mm}
  ~\\
  $\nearrow^*$\\
  ~
  \end{minipage}
  &
  \begin{minipage}[b]{5mm}
  \myfbox{$c$}\\[-0.290ex]
  \myfboxtwo{\raisebox{0.8ex}{$\vdots$}}
  \end{minipage}\\
  \setlength{\unitlength}{1mm}
    \begin{picture}(0,0)
    \put(-3.1,3.62){
      \put(0,0){\line(1,0){27.5}}
      \multiput(0,-4)(3,0){9}{\line(1,2){2}}}
  \end{picture}
  \end{array}
\]

\mypara{Running Time}

The \emph{number of configurations} that an automaton can enter during a computation depends on the input tape. The states and symbols are fixed for an automaton.
The number of head positions on the input tape is bounded by the \emph{length of the input tape}. The number of configurations is therefore linear in the length of the input tape, $n=O(|\mathit{tape}|)$. 
We remark that the number of configurations of an automaton with $k$ independent heads on the input tape is $n=O(|\mathit{tape}|^k)$. The $k$ head positions are easily accommodated by 
configurations of the form $c=(p,i_1,\ldots,i_k,A)$.
An automaton can carry out an \emph{exponential number of steps} before it terminates. For example, an automaton that during its computation forms all stacks consisting of $n$ zeros and ones takes $O(2^n)$ steps.

\mypara{Interpreter}

Figure~\ref{fig:dpdaint} shows the interpreter for {\iiDPDA} written in the style 
of an imperative language with recursion and call-by-value semantics. The interpreter \fc{\rawDint} can be run on a \emph{random-access machine} (RAM).
A call \fc{\Dint{c} = d} computes the terminator~\fc{d} of a configuration~\fc{c}, where \fc{\Apop{d}}. 
There is no symbol stack and no loop in the interpreter. All operations are modeled on the \emph{call stack} of the implementation language by recursive calls to the interpreter. A recursive call takes constant time, thus a call stack just adds a constant-time overhead compared to a data stack.
Statements \fc{\rawAacceptop} and \fc{\rawAhaltop} stop the interpreter and report whether the input was accepted or not. The automaton is assumed to be correct and no special
checks are performed by the interpreter.
We will now discuss the interpreter in more detail. It is the basis for the three interpreters and simulators in the following sections.

In the interpreter we abstract from the concrete push-, op- and pop-operations. %
We define predicates \fc{\Apush{c}}, \fc{\Aop{c}}, \fc{\Apop{c}}, \fc{\Aaccept{c}}, \fc{\Ahalt{c}} to be true if a configuration~\fc{c} causes the corresponding operation in the automaton.
Their actual effect on a configuration is not relevant as long as the next configuration can be determined by the built-in operations \fc{\rawAnext} and \fc{\rawAfollow}.
We let \mbox{\fc{\Anext{c}}} be the operation that yields in one step the next configuration, if \fc{\Aop{c}} or \fc{\Apush{c}}, and \mbox{\fc{\Afollow{c}{d}}} be the operation that yields in one step the next configuration given~\fc{c} and~\fc{d}, if \fc{\Apop{d}}.\footnote{The conventional `pop' just removes the top symbol from the stack. Our generalization that defines the next configuration by \fc{\Afollow{c}{d}} does not affect the complexity arguments later and is convenient from a programming language perspective.}
Each of these operations takes constant time, including \fc{\rawAnext} and \fc{\rawAfollow} that calculate the next configuration.

In case a configuration \fc{c} causes a pop-operation, that is \fc{\Apop{c}} is true in the cond-statement (Fig.~\ref{fig:dpdaint}), \fc{c} is a terminator and the interpreter returns it as result. If a configuration \fc{c} causes a push-operation, that is \fc{\Apush{c}} is true, first the terminator of the next configuration is calculated by \fc{\Dint{\Anext{c}} = d}. The terminator always causes a pop-operation and interpretation continues at configuration \fc{\Afollow{c}{d}} which follows from~\fc{c} and terminator~\fc{d}.
In case \fc{\Aop{c}} is true, that is the operation neither pushes nor pops, the terminator of \fc{c} is equal to the terminator \fc{\Dint{next(c)}} of the next configuration.

The effect of the operations on the configurations and the call stack can be summarized as follows.
\newcommand{\mycbox}[1]{\framebox[13mm]{\rule[-1mm]{0mm}{4mm}{#1}}}
\newcommand{\mydotbox}{
  \begin{picture}(37,0)
    \put(0,8.4){
      \put(0.2,0){\line(0,-1){11.5}}
      \put(0,-7.0){\makebox[13mm][c]{\fc{...}}}
      \put(36.8,0){\line(0,-1){11.5}}}
  \end{picture}}
\[
\begin{array}{r@{\hspace{8mm}}c@{\hspace{8mm}}l}
\\  \begin{minipage}[b]{23mm}
    \begin{tabular}[b]{r}
    \fc{c = }\mycbox{\fc{(p,i,A)}}\\[-0.29ex]
             \mydotbox
    \end{tabular}
    \end{minipage}
  & \begin{tabular}[b]{c}
      $\stackrel{\mbox{\fc{\Apush{c}}}}{\longrightarrow}$ \\[-0.29ex]
      ~
    \end{tabular}
  & \begin{minipage}[b]{40mm}
      \mycbox{\fc{(q,j,B)}}\fc{ = \Anext{c}}\\[-0.29ex]
      \mycbox{\fc{(p,i,A)}} \\[-0.29ex]
      \mydotbox
    \end{minipage}
  \\\\
    \begin{minipage}[b]{23mm}
    \begin{tabular}[b]{r}
      \fc{c = }\mycbox{\fc{(p,i,A)}}\\[-0.29ex]
               \mydotbox
     \end{tabular}
    \end{minipage}
  & \begin{tabular}[b]{r}
      $\stackrel{\mbox{\fc{\Aop{c}}}}{\longrightarrow}$\\[-0.29ex]
      ~
     \end{tabular}
  & \begin{minipage}[b]{40mm}
       \mycbox{\fc{(q,j,B)}}\fc{ = \Anext{c}}\\[-0.29ex]
       \mydotbox
    \end{minipage}
  \\\\
    \begin{minipage}[b]{23.26mm}
    \begin{tabular}[b]{r}
      \fc{d = }\mycbox{\fc{(q,j,B)}}\\[-0.10ex]
      \fc{c = }\mycbox{\fc{(p,i,A)}}\\[-0.29ex]
               \mydotbox
    \end{tabular}
    \end{minipage}
  & \begin{tabular}[b]{c}
      $\stackrel{\mbox{\fc{\Apop{d}}}}{\longrightarrow}$\\[-0.29ex]
      ~
     \end{tabular}
  & \begin{minipage}[b]{40mm}
      ~\\
      \mycbox{\fc{(r,k,C)}}\fc{ = \Afollow{c}{d}}\\[-0.29ex]
      \mydotbox
    \end{minipage}
  \\\\
\end{array}
\]
\begin{figure}[t]
\centering
\begin{minipage}[t]{115mm}
\begin{codesize}
\begin{alltt}\sffamily
\textbf{procedure} \Dint{c: conf}: conf; 
   \textbf{cond}
\begin{tabular}{l@{\ }l}
     \Apush{c}:   \myamp d := \Dint{\Afollow{c}{\Dint{\Anext{c}}}};\\
     \Aop{c}:     \myamp d := \Dint{\Anext{c}};\\
     \Apop{c}:    \myamp d := c;\\
     \Ahalt{c}:   \myamp \Ahaltop;\\
     \Aaccept{c}: \myamp \Aacceptop;
     \end{tabular}
   \textbf{end};
   \textbf{return} d
\end{alltt}
\end{codesize}
\end{minipage}
\mycaption{A recursive interpreter for deterministic {\PDA}.}
\label{fig:dpdaint}
\end{figure}%
A push-operation may, for example, push a constant symbol \fc{B} onto the stack or duplicate the current top~\fc{A}.
Likewise, an op-operation may replace the current top \fc{A} by a new top \fc{B}, but without pushing or popping the stack, and move the tape head by changing position~\fc{i} into~\fc{j}.
A pop-operation may just remove the stack top \fc{A} to uncover \fc{B} below or replace the uncovered symbol by a symbol \fc{C} depending on \fc{A} and \fc{B}. The abstract pop-operation covers many common binary stack-operations familiar from stack programming languages (\eg, it may choose from symbols \fc{A} and \fc{B} the one that is smaller according to some order). Depending on the concrete set of binary operators and stack symbols this allows to express a number of interesting functions as pushdown automata.

\mypara{Properties}

The body of the interpreter contains no loop, only sequential statements. The time it takes to execute each of the statements is bounded by a constant (ignoring the time to evaluate a recursive call to a result). No side-effects are performed and no additional storage is used except for the local variable~\fc{d}. Even though written in an imperative style, the interpreter is purely functional. It terminates if and only if the pushdown automaton terminates on the same input. The correctness of the interpreter should be evident as it merely interprets the automaton one step at a time. Note the simplicity of the construction by recursively calling the interpreter for each action of the automaton. Also an op-operation that does not change the height of the symbol stack converts into a (tail-recursive) call on the call stack.

In a terminating computation, no call \fc{\Dint{c}} can lead to a second call \fc{\Dint{c}} as long as the first call has not returned a result, which means that it is still on the call stack.
If a second call \fc{\Dint{c}} occurs while the first one is still on the call stack, the interpreter is in an infinite recursion.

As a consequence, in a terminating computation the height of the call stack is bounded by $n$, the number of configurations, and the same call stack cannot repeat during the interpretation. After exhausting all possible call stacks of height up to $n$, that is all permutations of up to $n$ configurations, the interpreter must terminate, that is within $O(n^n)$ steps. The interpreter can have a running time exponential in the number of configurations.

\section{Linear-Time Simulation of Deterministic {\PDA}}
\label{sec:Dsim}

The {\iiDPDA}-interpreter in Fig.~\ref{fig:dpdaint} is purely functional and has no persistent storage.
Each time the terminator~\fc{d} of a configuration~\fc{c} is computed and the same configuration is reached again, the terminator has to be recomputed by a call \fc{\Dint{c}}, which means the entire subcomputation is repeated. To store known terminators and to share them across different procedure invocations, we extend the interpreter with 
\emph{memoization}~\cite{Michie:68}.
This straightforward extension gives linear-time simulation of {\iiDPDA}. The sharing of terminators is the reason why Cook's speedup theorem works.

\begin{figure}[t]
\centering
\begin{minipage}[t]{115mm}
\begin{codesize}
\begin{alltt}\sffamily
\textbf{procedure} \Dsim{c: conf}: conf; 
   \textbf{if} \Adefined{T[c]} \textbf{then} \textbf{return} T[c];   /* \textit{find shortcut} */
   \textbf{cond}
\begin{tabular}{l@{\ }l}
     \Apush{c}:   \myamp d := \Dsim{\Afollow{c}{\Dsim{\Anext{c}}}};\\
     \Aop{c}:     \myamp d := \Dsim{\Anext{c}};\\
     \Apop{c}:    \myamp d := c;\\
     \Ahalt{c}:   \myamp \Ahaltop;\\
     \Aaccept{c}: \myamp \Aacceptop;
     \end{tabular}
   \textbf{end};
   T[c] := d;                                 /* \textit{memoize result} */
   \textbf{return} d
\end{alltt}
\end{codesize}
\end{minipage}
\mycaption{A linear-time simulator for deterministic {\PDA}.}
\label{fig:dpdasim}
\end{figure}

\mypara{RAM extension}
Figure~\ref{fig:dpdasim} shows the interpreter with memoization, called \emph{simulator}.
It works in the same way as the interpreter except that
each time before a call \fc{\Dsim{c}} returns the terminator \fc{d} of \fc{c}, the terminator is stored in a table \fc{T} by assignment \fc{T[c] := d}. Next time the terminator is needed, it can be retrieved from~\fc{T}, 
avoiding its recomputation. Terminators are now shared dynamically at run time and over the entire simulation. Table \fc{T} can be implemented as a one-dimensional array indexed by configurations and can hold one terminator for each of the $n$ configurations that can occur during a computation. All table entries are initially undefined. It is easy to see that the shortcut (if-statement) and the memoization (table assignment) do not change the result of the automaton. Storing and retrieving a terminator takes constant time on a RAM (see Cook for a charged RAM model instead of a unit-cost model~\cite{Cook:72}). 
An ``automatic storage management'' also means that many terminators are recorded during a computation that are not needed later, but we shall see that this does not affect the linearity argument. A more thorough analysis would surely reveal that memoization points are only required at a few places in an automaton (\cf~\cite{AJ:94,Mogensen:94}).

\mypara{Linear-time simulation} In a terminating computation, before a second call \fc{\Dsim{c}} is made, the first call must have returned and stored the terminator~\fc{d} of~\fc{c} at \fc{T[c]}. Once the terminator is known, it need not be recomputed and can be fetched from the table.
Hence, the cond-statement,
which is guarded by a lookup in \fc{T}, is executed \emph{at most once} for any \fc{c}.
Recursive calls to the simulator occur only from within the cond-statement, namely one call if \fc{\Aop{c}} and two calls if \fc{\Apush{c}}. Consequently, \fc{\rawDsim} can be called at most $2n$ times, where $n$ is the number of possible configurations. This also limits how often the if-statement guarding the cond-statement is executed. Hence, the total number of execution steps during a terminating simulation is bounded linearly by~$n$.
Recall that $n$ is linear in the length of the input tape, $n=O(|\mathit{tape}|)$. This concludes the argument for the linear-time simulation of a {\iiDPDA} on a~RAM.

\mypara{Discussion} Deterministic pushdown automata are the accepting device for \emph{deterministic context-free languages}. More precisely, they are exactly recognized by \emph{1-way} deterministic pushdown automata ({\iDPDA}), that is,  deterministic pushdown automata that never move their head to the left on the input. The LR grammar of a deterministic context-free language is easy to convert into a {\iDPDA} (\eg~\cite{HopcroftUllman:79:book}). 
Thus, recognition of this subclass of context-free languages using the memoizing simulator \fc{\rawDsim} (Fig.~\ref{fig:dpdasim}) takes at most linear time (as does the classic LR-parsing algorithm by Knuth).
In the following we extend the simulator to recognize all context-free languages in cubic time.

The method by Aho {\etal}~\cite{AHU:68} requires $O(n^2)$ for simulating {\iiDPDA}, a result which was then strengthened to $O(n)$ by Cook~\cite{Cook:72}. Both methods work bottom-up. In contrast, the simulator \fc{\rawDsim}  works top-down following the forward control flow
as does the one by Jones~\cite{Jones:77}. It clearly shows that the key mechanism that turns a recursive pushdown interpreter into a linear-time simulator is 
memoization.

\section{Interpretation of Nondeterministic {\PDA}}
\label{sec:Nint}

In a two-way \emph{nondeterministic} pushdown automaton ({\iiNPDA}) the computation path is not uniquely determined. A deterministic automaton can be made \emph{nondeterministic} by introducing an operation {\bf choose} that allows the automaton to select any of two computation paths in a configuration \fc{c} (\cf~\cite{Jones:97:complexity}). This means that a configuration no longer has a unique terminator, but a set of possible terminators.
We let \fc{\Anextl{c}} and \fc{\Anextr{c}} be the abstract operations that yield in one step the two next configurations that are possible if \fc{\Achoose{c}}.
For simplicity, the new operation can neither push nor pop stack symbols.
With a choose-operation two transitions are possible:
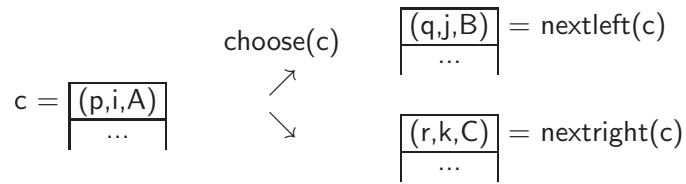
\[
\begin{array}{l@{\hspace{7mm}}c@{\hspace{8mm}}l}
  \begin{minipage}[b]{23mm}
    \begin{tabular}[t]{r}
    \fc{c = }\mycbox{\fc{(p,i,A)}}\\[-0.29ex]
    \mydotbox
    \end{tabular}
  \end{minipage}
  & \raisebox{-1.8ex}{\begin{minipage}[b]{15.5mm}
      \centering
      \fc{\Achoose{c}}\\[0.3ex]
      $\nearrow$\\[0.5ex]
      $\searrow$
    \end{minipage}}
  &
  \begin{minipage}[c]{40mm}
    \mycbox{\fc{(q,j,B)}}\fc{ = \Anextl{c}}\\[-0.29ex]
    \mydotbox\\[3.0ex]
    \mycbox{\fc{(r,k,C)}}\fc{ = \Anextr{c}}\\[-0.29ex]
    \mydotbox\\[-1.1ex]
  \end{minipage}
\end{array}
\]
A nondeterministic automaton is said to \emph{accept} an input if it has at least one accepting computation when started in the initial state $p_0$ with an empty stack and the head at the left tape end. It has the ability to guess the right choice that leads to the shortest accepting computation. In an interpreter this ``angelic nondeterminism'' can be thought of as searching through a tree of all possible computation sequences, some of which may be infinite or non-accepting,
to find at least one accepting sequence. Branching in the computation tree is due to nondeterministic choose-operations in the automaton.

\mypara{Interpreter}

\begin{figure}[t]
\centering
\begin{minipage}[t]{115mm}
\begin{codesize}
\begin{alltt}\sffamily
\textbf{procedure} \Nint{c: conf}: confset;
   \textbf{if} \Avisited{T[c]} \textbf{then} \textbf{return} \{\ensuremath{\,}\};      \:/* \textit{detect infinite branch} */
   T[c] := Visited;                           /* \textit{mark configuration} */
   \textbf{cond}
\begin{tabular}{l@{\ }l}
     \Apush{c}:   \myamp d := \myma{\bigcup}\:\Nint{\Afollow{c}{e}} \textbf{where} e \myma{\in} \Nint{\Anext{c}};\\
     \Aop{c}:     \myamp d := \Nint{\Anext{c}};\\
     \Achoose{c}: \myamp d := \Nint{\Anextl{c}} \myma{\cup} \Nint{\Anextr{c}};\\
     \Apop{c}:    \myamp d := \{\ensuremath{\,}c\ensuremath{\,}\};\\
     \Ahalt{c}:   \myamp d := \{\ensuremath{\,}\};\\
     \Aaccept{c}: \myamp \Aacceptop;
   \end{tabular}
   \textbf{end};
   T[c] := Undef;                            /* \textit{unmark configuration} */ 
   \textbf{return} d
\end{alltt}
\end{codesize}
\end{minipage}
\mycaption{A recursive interpreter for nondeterministic {\PDA}.}
\label{fig:npdaint}
\end{figure}

The interpreter for \emph{nondeterministic} {\PDA} that can be run on a RAM is shown in Fig.~\ref{fig:npdaint}.
Two main changes to the original interpreter in Fig.~\ref{fig:dpdaint} are necessary to accommodate the ``guessing'':
(1)~a \emph{set of terminators} instead of a single terminator is returned, and
(2)~a \emph{termination check} (``seen before'') that stops interpretation along an infinite computation sequence. We detail the two modifications below.
\begin{enumerate}
\item
\emph{Terminator sets}:
A choose-operation requires the union of the terminator sets obtained from the two next configurations, \fc{\Anextl{c}} and \fc{\Anextr{c}}. In case of a push-operation, and this is the most involved modification, each configuration \fc{e} in the terminator set obtained by the inner call \fc{\Nint{\Anext{c}}} must be followed by an outer call. The big set union used for this purpose is a shorthand for a while-loop over the inner terminator set. A pop-operation now returns a singleton set \fc{\{\ensuremath{\,}c\ensuremath{\,}\}} instead of~\fc{c}. Finally, instead of making a full stop at a halt-operation, an empty set is returned in order not to miss an accepting computation along another possible branch.

\item

\emph{Termination check}: As discussed before, non-termination occurs when the interpreter is called a second time with the same configuration \fc{c} as argument while the first call has not yet returned.
This situation can be detected by marking \fc{c} in a table when a call \fc{\Nint{c}} is made and unmarking \fc{c} when the call returns. If a call with a marked \fc{c} as argument is made, an infinite computation is detected and the interpreter returns an empty terminator set. The same table \fc{T} as before can be used, but can now hold the additional value \fc{Visited}. Initially all table entries are set to \fc{Undef}.
\end{enumerate}
The cardinality of a terminator set is bounded by $n$, the number of configurations that can occur in a computation. The most costly set operation in the interpreter is the union of terminator sets. Assuming a suitable choice of data structures, a union operation takes time linear in the total cardinality of the sets, that is the union of two sets with cardinalities $u$ and $v$ takes time $O(u+v)$. All remaining set-operations needed in the interpreter are straightforward and take constant time: creating a set (empty, singleton), and picking and removing an arbitrary element from a set (in the set comprehension). In the discussion below we assume 
such an implementation of the set operations.%
\footnote{A straightforward implementation of such a set data structure might be a Boolean array of length $n$ to indicate membership of a configuration~\fc{c} in a set together with an unsorted list of all configurations contained in that set.}

A choose-operation, which unites two terminator sets each of cardinality up to $n$, takes linear time $O(n)$. A push-operation,
where the inner call \fc{\Nint{\Anext{c}}} returns a set of at most $n$ terminators, each of which, when followed by the outer call \fc{\Nint{\Afollow{c}{e}}}, can again return a set of up to $n$ terminators, requires the union of $n$ sets each of cardinality up to $n$, which then takes quadratic time $O(n^2)$. This is the most expensive set-operation in the cond-statement.

In the case of a deterministic automaton, that is, an automaton \emph{without} choose-operation, the new interpreter in Fig.~\ref{fig:npdaint} operates with singleton sets only, and the set-operations introduce at most a constant-time overhead compared to the original interpreter in Fig.~\ref{fig:dpdaint}. This is useful because the new interpreter ``falls back'' to its original behavior and, except for a constant time overhead in the new interpreter, there is no penalty in using it to run deterministic {\PDA} and, as an extra benefit, it always terminates.

There is a major pitfall. If a nondeterministic automaton is left-recursive, then the termination check may stop left-recursion too early and miss useful branches contributing to a terminator set. In the case of {\iNPDA} there always exists a non-left-recursive version (presumably the same for {\iiNPDA}). Alternatively, one might bound the unfolding of a left-recursion in terms of the input assuming some normal-form automaton (the termination check in Fig.~\ref{fig:npdaint} limits left-recursion unfolding to one).

\section{Cubic-Time Simulation of Nondeterministic {\PDA}}
\label{sec:Nsim}

\begin{figure}[t]
\centering
\begin{minipage}[t]{115mm}
\begin{codesize}
\begin{alltt}\sffamily
\textbf{procedure} \Nsim{c: conf}: confset; 
   \textbf{if} \Adefined{T[c]} \textbf{then} \textbf{return} T[c];   /* \textit{find shortcut } */
   \textbf{if} \Avisited{\ensuremath{\,\;}T[c]} \textbf{then} \textbf{return} \{\ensuremath{\,}\};    \:/* \textit{detect infinite branch} */
   T[c] := Visited;                          /* \textit{mark configuration} */
   \textbf{cond}
\begin{tabular}{l@{\ }l}
     \Apush{c}:   \myamp d := \myma{\bigcup}\:\Nsim{\Afollow{c}{e}} \textbf{where} e \myma{\in} \Nsim{\Anext{c}};\\
     \Aop{c}:     \myamp d := \Nsim{\Anext{c}};\\
     \Achoose{c}: \myamp d := \Nsim{\Anextl{c}} \myma{\cup} \Nsim{\Anextr{c}};\\
     \Apop{c}:    \myamp d := \{\ensuremath{\,}c\ensuremath{\,}\};\\
     \Ahalt{c}:   \myamp d := \{\ensuremath{\,}\};\\
     \Aaccept{c}: \myamp \Aacceptop;
   \end{tabular}
   \textbf{end};
   T[c] := d;                                 /* \textit{memoize result} */
   \textbf{return} d
\end{alltt}
\end{codesize}
\end{minipage}
\mycaption{A cubic-time simulator for nondeterministic {\PDA}.}
\label{fig:npdasim}
\end{figure}

To turn the new interpreter (Fig.~\ref{fig:npdaint}) into a fast simulator (Fig.~\ref{fig:npdasim}) we use the same memoization method as in Sect.~\ref{sec:Dsim}. The use of table \fc{T} parallels its use in the deterministic case except that for each of the $n$ possible configurations the table can now hold a set of up to $n$ terminators and the value \fc{Visited}.
The body of the simulator is again guarded by an if-statement (first line) that returns the terminator set of a configuration \fc{c}, if it is available in table \fc{T}. Otherwise, and if no infinite computation path is detected, \fc{c} is marked as \fc{Visited} in \fc{T} and its terminator set is computed.

Before returning, terminator set~\fc{d} of~\fc{c} is stored in~\fc{T}, which overwrites the mark \fc{Visited}.
The cond-statement is executed at most once for each configuration. The mark \fc{Visited} is only needed the first time the procedure is called, when the table  does not yet contain a terminator set for~\fc{c}. Thus, the same table can be used for marking configurations and for storing their terminator sets. A terminator set may be empty if none of the branches rooted in \fc{c} is accepting. Otherwise, the interpreter is unchanged.

\mypara{Cubic-time simulation}

As before, the cond-statement is executed at most once for each of the $n$ configuration due to the guards at the beginning of \fc{\rawNsim}. Up to $n+1$ calls to \fc{\rawNsim} may occur in the case of a push-operation, namely one inner call and at most $n$ outer calls, one for each \fc{e $\in$ \Nsim{\Anext{c}}}. Hence, \fc{\rawNsim} can be called at most $O(n^2)$ times during a simulation. This also limits how often the if-statements guarding the cond-statement are executed.

In the cond-statement, as before, the simulation of the op-, pop-, halt-, accept-operations takes constant time, $O(1)$. The union of two sets of at most $n$ terminators in case of a choose-operation may take linear time, $O(n)$. 
The union of the terminator sets in a push-operation is the most costly operation and may take quadratic time, $O(n^2)$.
A push is simulated at most once per execution of a cond-statement, which is at most $n$ times.
Hence, the total number of execution steps during a simulation is cubic in the number of configurations, $O(n^3)$. Recall that $n$ is linear in the length of the input tape, $n=O(|\mathit{tape}|)$. This ends the argument for the
cubic-time 
simulation of (non-left-recursive) {\iiNPDA} on a RAM.

\mypara{Discussion}

We observe that the ``complexity generator'' in the cond-statement is \emph{not} the choose-operation, even though it introduces two computation branches, rather the handling of up to $n$ continuations and the union of their terminator sets in case of a push-operation. If the cardinality of each terminator set that can occur during a simulation is bounded by a \emph{constant}, that is, not dependent on the input, the simulation time is \emph{linear} in the input as before. 
Deterministic automata, where the cardinality of each terminator set is at most one, and a class of nondeterministic automata, 
where the cardinality is bounded by some $k$,
are all simulated in linear time by \fc{\rawNsim}. The top-down method is useful because the same simulator runs them 
in the time corresponding to their degree of nondeterminism.

\emph{One-way} nondeterministic pushdown automata ({\iNPDA}) are the accepting device for \emph{context-free languages}. Every context-free language has a grammar without left recursion and it is straightforward to convert the grammar into a {\iNPDA}. This means that recognition of context-free languages using the simulator (Fig.~\ref{fig:npdasim}) has the same worst-case time complexity as the classic parsing algorithms that can handle the full class of context-free languages (Earley, Cocke-Younger-Kasami), 
that is $O(|\mathit{string}|^3)$. As discussed before, the performance of the simulator is determined by the degree of nondeterminism in the automaton. Recognition of deterministic context-free languages using the simulator takes, again, at most linear time. In practice, of course, specialized parsing algorithms will have better run times (due to the constant term hidden in the $O$-notation) and use less space than the recursive simulator.
Again, the mechanism that enables polynomial-time simulation is the sharing of computations by memoization.

\mypara{Acknowledgements}

%Thanks 
%for various useful comments %
%are due to
%Nils Andersen,
%Holger Bock Axelsen,
%Julia Lawall,
%Torben Mogensen, and
%Chung-chieh Shan, and to
%Neil D.\ Jones for pointing the author to 
%%
%Cook's construction.
%%

Thanks are due to 
Nils Andersen,
Holger Bock Axelsen,
Julia Lawall,
Torben Mogensen,
Chung-chieh Shan, 
and the anonymous reviewers for various insightful comments,
to
Neil D.\ Jones for pointing the author to 
Cook's construction, and to Akihiko Takano for providing the author with excellent working conditions at the National Institute of Informatics, Tokyo.

\bibliographystyle{eptcs}
\bibliography{pemc,russian}

\end{document}